\documentclass[twocolumn,preprintnumbers,superscriptaddress,nofootinbib,aps,prl,floatfix]{revtex4}

\usepackage{amsmath,amssymb}
\usepackage{dsfont}
\usepackage{graphicx} 
\usepackage{epstopdf} 
\usepackage{slashed}
\usepackage{subfigure}
\usepackage{color}
\usepackage{multirow}

\usepackage{hyperref}

\usepackage{footmisc}

\usepackage{array}
\newcolumntype{L}[1]{>{\raggedright\let\newline\\\arraybackslash\hspace{0pt}}m{#1}}
\newcolumntype{C}[1]{>{\centering\let\newline\\\arraybackslash\hspace{0pt}}m{#1}}
\newcolumntype{R}[1]{>{\raggedleft\let\newline\\\arraybackslash\hspace{0pt}}m{#1}}

\newcommand{\be}{\begin{eqnarray*}}
\newcommand{\ee}{\end{eqnarray*}}

\newcommand{\bee}{\begin{eqnarray}}
\newcommand{\eee}{\end{eqnarray}}
\newcommand{\beeq}{\begin{equation}}
\newcommand{\eeeq}{\end{equation}}

\usepackage{xspace}

\def\pT{\ensuremath{p_{\mathrm{T}}}}
\def\antibar#1{\ensuremath{#1\bar{#1}}}
\def\sherpa{\textsc{Sherpa}\xspace}
\def\pythia{\textsc{Pythia}\xspace}
\def\MadGraph{\texttt{Madgraph}\xspace}

\def\powhegbox{\textsc{Powheg-Box}\xspace}
\def\MET{\ensuremath{E_{\mathrm{T}}^{\mathrm{miss}}}}

\begin{document}

\title{Reconstructing singly produced top partners in decays to $\mathbf{Wb}$}

\begin{abstract}
  Fermionic top partners are a feature of many models of physics beyond the Standard Model.  We propose a search strategy for single production of top partners focussing specifically on the dominant decay to $Wb$. The enormous background can be reduced by exploiting jet substructure to suppress top-pair production and by requiring a forward jet. This simple strategy is shown to produce a sensitive search for single top-partner production, in the context of composite Higgs models, that has competitive mass reach with existing experimental searches for top-partner-pair production at the 8\,TeV LHC.
\end{abstract}
\author{Nicolas Gutierrez} 
\affiliation{School of Physics and Astronomy, University of
  Glasgow,\\Glasgow, G12 8QQ, United Kingdom\\[0.1cm]}
\author{James Ferrando} 
\affiliation{School of Physics and Astronomy, University of
  Glasgow,\\Glasgow, G12 8QQ, United Kingdom\\[0.1cm]}
\author{Deepak Kar  } 
\affiliation{School of Physics and Astronomy, University of
  Glasgow,\\Glasgow, G12 8QQ, United Kingdom\\[0.1cm]}

\author{Michael Spannowsky} 
\affiliation{Institute for Particle Physics Phenomenology, Department
  of Physics,\\Durham University, Durham, DH1 3LE, United Kingdom\\[0.1cm]}

\pacs{}
\preprint{IPPP/14/25, GLAS-PPE/2014-01,  DCPT/14/50}
\maketitle


\section{Context and Introductory Remarks}

With the observation of a Standard Model-like Higgs boson~\cite{Aad:2012tfa,Chatrchyan:2012ufa} at the Large Hadron Collider (LHC), 
our understanding of electroweak symmetry breaking has been significantly enhanced. The 
attention of experiments at the LHC now turns to establishing the properties 
of this new resonance. The possibility that the resonance is actually a 
composite bound state remains open, and in this work we study a top partner search in the context of
 composite Higgs scenarios~\cite{Kaplan:1983fs,Kaplan:1983sm,Georgi:1984af,Dugan:1984hq,Contino:2003ve,Agashe:2004rs,Contino:2006nn}. The discovery of such top partners would spectacularly elucidate the 
mechanism by which the scalar resonance mass is stabilized at the electroweak scale.

Although the Higgs boson can be a generic composite bound state, we focus on the realization of the Higgs boson as a pseudo Nambu-Goldstone boson (pNGB) of the coset $G/H$, $G$ is an approximate global symmetry and $H$ is an unbroken subgroup. 
We assume partial compositeness \cite{Kaplan:1983fs} to evade flavour physics constraints~\cite{Kaplan:1991dc}. Here, the Standard Model (SM) fermions obtain masses via mixing with  composite bound states from the strongly coupled sector. 
Since SM fermions arise as admixtures of the elementary and the corresponding composite bound states, there are necessarily accompanying heavy excitations with the same SM quantum numbers, so-called top partners. They belong to a representation of the unbroken subgroup $H$.

Following Ref.~\cite{DeSimone:2012fs} we adopt a minimal setup for $G=SO(5) \times U(1)_X$ and $H=SO(4) \times U(1)_X$. We assume $t_R$ to be a completely composite chiral state and only $q_L = (b,t)_L$ to be partially composite. In this short article we focus on evaluating the LHC sensitivity for discovering or excluding a top partner belonging to the singlet $\Psi = {\bf 1}_{2/3}$ of the unbroken SO(4). The operators $\mathcal{O}$ induced by the interactions of the fermions in the strong sector are assumed to transform in the representation $r_\mathcal{O}=5_{2/3}.$\footnote{Our results can be interpreted straightforwardly if the operators $\mathcal{O}$ of the strong-sector fermions transform in the representation $r_{\mathcal{O}}={\bf 14}_{2/3}$.}

The interaction Lagrangian for $\tilde{T}$ is given by

\begin{eqnarray}
\mathcal{L} &=& \mathcal{L}_{\mathrm{kin}} - M_{\Psi} \bar{\Psi} \Psi \nonumber \\
&+& \left [ y f (\bar{Q}^5_L)^I U_{I5} \Psi_R + y c_2 f (\bar{Q}^5_L)^I U_{I5} t_R + \mathrm{h.c.} \right ]
\end{eqnarray}
and the mass eigenvalue of the top partner $\tilde{T}$ is approximately given by
\begin{equation}
m_{\tilde{T}} \simeq M_\Psi \left [ 1 + \frac{y^2}{4 g^2_\Psi} \frac{v^2}{f^2}  \right ],
\end{equation}
where $v=246~\mathrm{GeV}$ and $g_\Psi = M_\Psi / f$. 

As a result, the phenomenology of $\tilde{T}$ is determined by four parameters: $(M_\Psi, y, c_2, f)$. $M_\Psi$ is the mass of the top partner, $y$ controls the mixing between the composite and elementary states, $c_2$ is an $\mathcal{O}(1)$ parameter associated with the interactions of $t_R$ and $f$ is the symmetry breaking scale of the strong sector. By requiring the four parameters to conspire to give the observed top quark mass, one degree of freedom in this parameter space can be removed, i.e. we choose $y=1$.
In the following quantitative analysis we also choose $\xi=\frac{v^2}{f^2}=0.2$ to ensure compatibility with electroweak precision tests \cite{Barbieri:2007bh,Giudice:2007fh,Grojean:2013qca}. The cross section also depends on the parameter $c_2$, as described in Ref~\cite{DeSimone:2012fs}. Here we choose $c_2=0.891767$ for a $\tilde{T}$ mass of 700\,GeV.

Pair production of fermionic top partners has been thoroughly searched for at the LHC~\cite{ATLAS:2012qe,Chatrchyan:2013uxa}.
In composite Higgs models, light top partners are well motivated \cite{Matsedonskyi:2012ym} but so far masses less than 687 -- 782\,GeV have been excluded depending on the branching ratio to
 different final states. In this work, we focus instead on single production: $p p \rightarrow q\tilde{T}b$. The cross section 
for single production is smaller  than that for pair production at low masses of the top partner.
 However, depending on the  weak coupling to the top partner, single production can begin to have a
 larger cross section in the mass-range of 600--1000\,GeV~\cite{Aguilar-Saavedra:2013qpa}. Thus, 
well-designed searches for single production can potentially extend the mass-reach 
of the LHC experiments.

Singlet top partners decay to $Wb$,$tZ$ or $tH$ in the approximate ratio 2:1:1 respectively, and the significant
branching fractions to $tZ$ and $tH$ have attracted previous attention\cite{Mrazek:2009yu, DeSimone:2012fs, Vignaroli:2012nf}. In particular, multi-leptonic
final states arising from these decays make for searches with low background, albeit at a 
relatively low overall signal efficiency. Here, we have chosen to focus on the most abundant expected  decay: $\tilde{T} \rightarrow Wb$, with subsequent decay of the $W$ boson to leptons \footnote{This final state could also be used to search for exotic bottom partners~\cite{Alvarez:2013qwa}, although we do not consider that possibility here.}. Trading a clean final state for a larger signal efficiency can be beneficial for large top partner masses \cite{Azatov:2013hya}. The present experimental limits are weakest for large values of the $\tilde{T}\rightarrow Wb$ branching ratio~\cite{Chatrchyan:2013uxa}. We demonstrate that, by requiring a forward 
jet and exploiting jet substructure to suppress backgrounds from top-quark production, this relatively neglected channel can become a promising one for discovery of single top-partner production at the LHC.

\section{Elements of the analysis}

\begin{figure*}[htb]
  \includegraphics[width=0.70\textwidth]{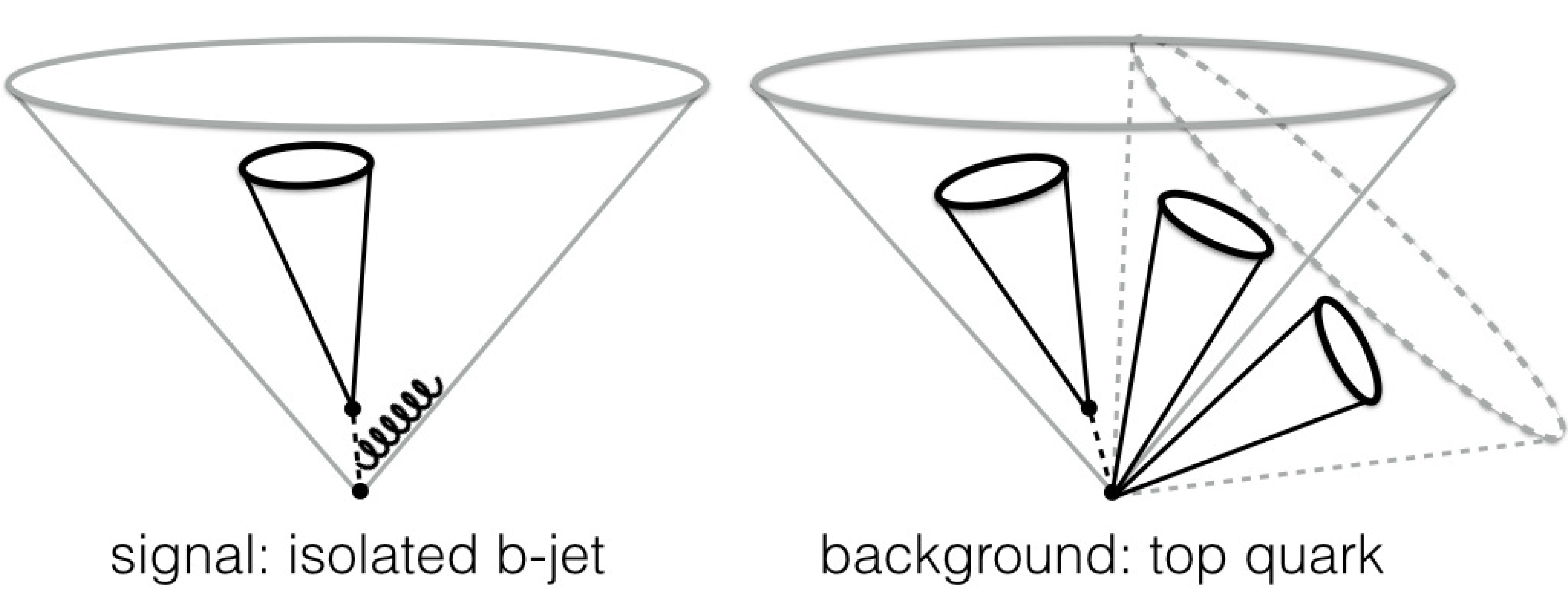}
  \caption{Illustration of the usage of large-radius jet mass to veto $t\bar{t}$ background. For the signal $\tilde{T} \rightarrow Wb$, the $b$-quark recoils against the $W$-boson. Thus the hardest large-radius jet in the event typically contains a $b$-hadron plus additional soft and collinear radiation, and tends to have  a low mass.
	         For the semi-leptonic $t\bar{t}$ background, right, a mildly boosted hadronically decaying top quark produces large-radius jets containing a significant fraction of the top decay products. The fraction of top decay products contained, and therefore the jet mass, increases with jet \pT.
	         Hence, after requiring a matching between a small-radius $b$-jet  and  large-radius jet,  a cut based on the large-radius jet \pT\ and mass can be optimized to distinguish between signal and $t\bar{t}$ background, whilst still retaining good signal efficiency.
	         }
  \label{fig:Illustration}
\end{figure*}

\begin{figure}[!b]
  \includegraphics[width=0.43\textwidth]{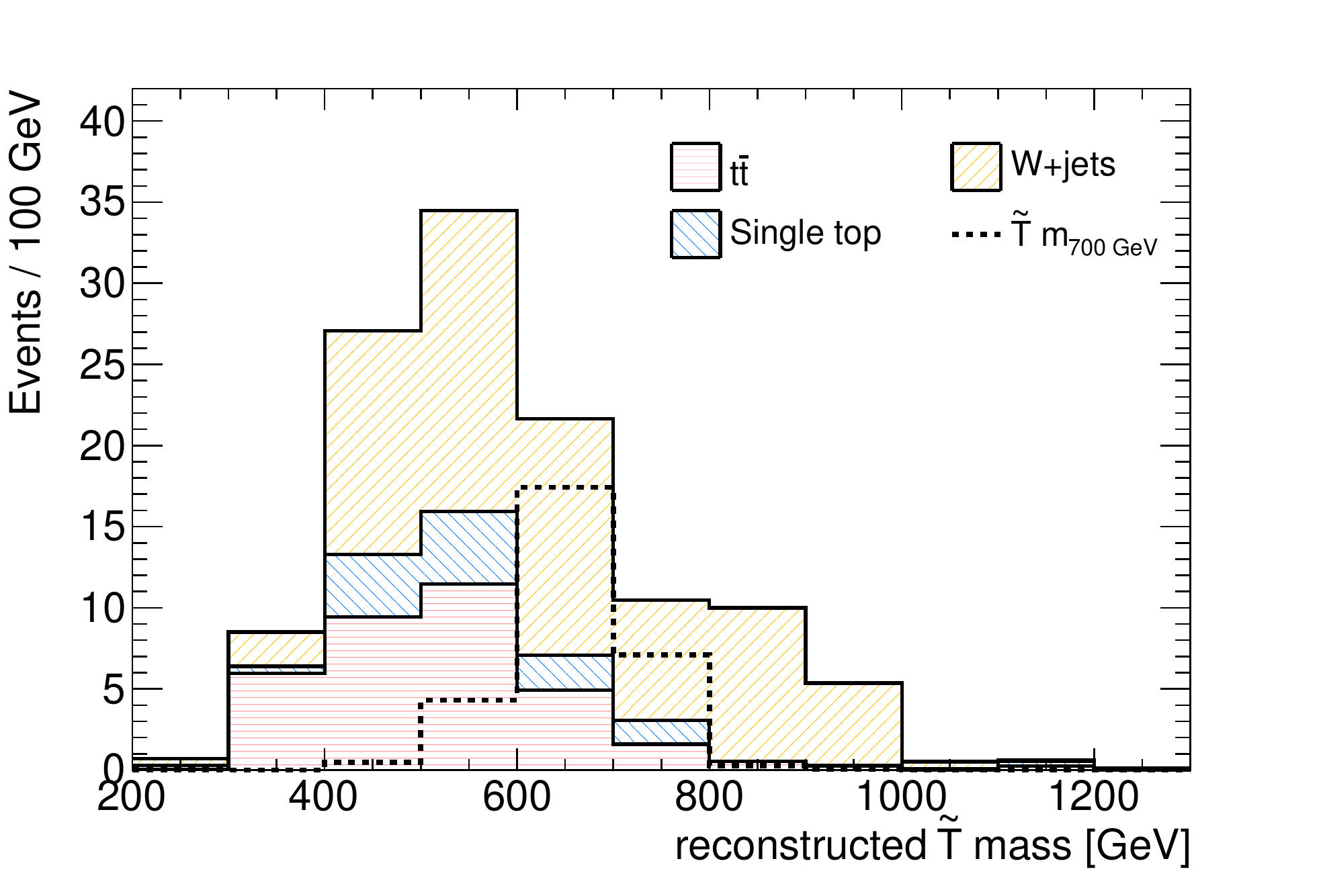}
  \caption{The distribution of the $\tilde{T}$ candidate mass used in the statistical analysis is shown
     for 20\,fb$^{-1}$ integrated luminosity at the 8\,TeV LHC. 
     The shape of the SM background from $W$+jets, $t\bar{t}$ and single top production are compared to the shape of the expected signal.}
  \label{fig:Tmass} 
\end{figure}

\subsection{Samples}

Signal events were simulated using \MadGraph5~\cite{Alwall:2011uj} interfaced with \pythia 8.185~\cite{Sjostrand:2007gs} for parton-shower and fragmentation. For a top partner with mass 700\,GeV, the leading-order cross sections from \MadGraph are 0.163\,pb (0.957\,pb) for $\sqrt{s}=8$\,TeV (14\,TeV)\footnote{The branching ratio $\tilde{T}\rightarrow Wb$ for our parameter choices is 0.55}.
Background samples of \antibar{t} and $W$+jets were generated with up to two and four additional partons using \sherpa~\cite{Gleisberg:2008ta} 2.1.0.
Single top quark production in the $t$-channel
was modelled, in a four-flavor scheme~\cite{singtop-tchan-4f}, using \powhegbox~\cite{powhegbox} showered with \pythia 8.185.
   
The \antibar{t} production cross section of the \sherpa sample was scaled
to the next-to-next-to-leading-order prediction including resummation  of next-to-next-to-leading logarithmic soft gluon terms~\cite{toppred} calculated using the NNPDF2.3 PDF set~\cite{NNPDF}. Approximate corrections to next-to-leading order for the signal cross-section were calculated using the single-top cross section from \powhegbox, with the top-mass set to the mass of the top-partner compared to the equivalent leading order cross-section calculated from \MadGraph. 
These corrections to the signal were approximately 14(7)\% for 8(14)\,TeV.

\subsection{Event selection and top-partner reconstruction}

Analysis of the samples of simulated events is performed on the stable final-state particles using {\tt Rivet}~2.1.1~\cite{Rivet}. The event selection was designed to select singly produced  $\tilde{T}$ quarks with subsequent $ \tilde{T} \rightarrow Wb$ decay. 
For leptonically decaying $W$-bosons, the signal event topology is characterized by a charged lepton, missing transverse momentum, a high-\pT\  $b$-tagged central jet and a forward jet.

 Final-state electrons are corrected for energy loss due to photon emission by combining them with all final-state photons within  $\Delta R(e,\gamma)  < 0.1$, where $\Delta R(e,\gamma)$ is the distance in the $\eta$--$\phi$ plane between the electron and the photon. Charged leptons are required to be isolated from other particles\footnote{Isolation criteria that were similar to those used in an ATLAS same-sign dilepton search~\cite{ATLAS:2012mn} were adopted. For electrons, the isolation requirement is that the total \pT\ of all charged particles ($p_q$) , $\sum$\pT, with $\Delta R(e,p_q)<0.3$ should be less than 10\%  of the electron \pT. Similarly for muons, $\sum$\pT\ for all particles with $\Delta R(\mu,p)<0.4$ is required to be less than 6\%
of the muon \pT, but in addition, $\sum$\pT\ is required to be less than 
$4+$ \pT\ $\times 0.02$.}.
The typical fiducial acceptance of the ATLAS detector~\cite{Collaboration:2010knc} is mimicked by only using electrons with $1.52 < |\eta| < 2.47$ or $|\eta|<1.37$, and muons with $|\eta| < 2.5$.

Small-radius jets are clustered from all final-state particles with $|\eta| < 5.0$, except muons and neutrinos,
using the anti-$k_{t}$ algorithm~\cite{Cacciari:2008gp,Cacciari:2011ma} with a radius parameter of 0.4. Only jets with \pT$\geq$ 25\,GeV for $|\eta|<2.4$
or \pT$\geq$ 35\,GeV for $2.5 < |\eta| < 4.5$ are used.
In order to account for electrons misidentified as small-radius jets, any  small-radius jet ($j$) with $\Delta R (e,j) \leq 0.2$ are discarded.
Large-radius jets are clustered in a similar manner, but with a radius parameter of 1.0. These large-radius jets were required to have \pT$\geq$ 100 GeV and $|\eta|<$1.2.

Jets are referred to as $b$-jets if their constituents contain a bottom flavoured hadron or the decay products of at least one such hadron. The typical performance of experimental jet 
$b$-tagging was mimicked by ascribing a 0.7 probability for tagging $b$-jets and a 0.01 mis-tag
rate for $b$-tagging jets from other sources~\cite{Aad:2013nca}.

The missing transverse momentum, \MET\ , is the magnitude of the total transverse momentum calculated from all final-state particles with $|\eta|<5.0$ except neutrinos. The missing transverse momentum direction is defined as being directly opposite to this same total transverse momentum.

Events are preselected by requiring exactly one electron or muon with \pT\ $\geq$ 25\,GeV, \MET $\geq$ 20\,GeV and \MET\  plus transverse mass\footnote{We define the transverse mass of the lepton and \MET to be $m_{\mathrm{T}}= \sqrt{2\pT^{l} \MET(1-\cos \phi_{l\nu}})$, where $\cos \phi_{l\nu}$ is the azimuthal angle between the charged lepton and the missing transverse momentum direction.} of 60\,GeV.
Events are also required to contain at least two small-radius jets and  at least one large-radius jet with \pT\ $> 200$\,GeV.  After this selection, the dominant background processes are $W$+jet 
production and $t\bar{t}$ production. Further cuts were applied with the goal of optimizing the
 statistical significance $S/ \sqrt{B}$, defined as the number of 700\,GeV top-partner signal ($S$)  events divided by the square root of the number of background ($B$) events. 

The $W+$ jet background is suppressed by requiring that a $b$-tagged small-radius jet is matched to the large-radius jet\footnote{Small radius jets are considered matched to large radius jets if their four vectors lie within an $\eta$--$\phi$ distance of 0.8.}. The $t\bar{t}$ background is also suppressed by requiring that the large-radius jet have $\pT\ > 250$\,GeV and mass $< 70$\,GeV. The latter requirement exploits the fact that high $p_T$ large-radius jets in $t\bar{t}$ tend to have a genuinely large mass since they contain at least two decay products from $t\rightarrow Wb \rightarrow \bar{q} q'b$, as illustrated in Figure~\ref{fig:Illustration}. To ensure that the lepton is not within the active area of the large-radius jet, it is required that $\Delta \phi (l,\mathrm{jet})> 1.5$.

The expected isolation of the b-tagged jet from the $\tilde{T}$ decay can be exploited further by vetoing events with extra central jets (those with $|\eta| < 2.4$) above a certain \pT\ threshold.
Theoretical uncertainties when vetoing extra jets decrease for higher thresholds, i.e. the resummation of contributions $\sim \ln^2 (\sqrt{\hat{s}}/p_{T,\mathrm{veto}})$ becomes increasingly important for lower cut-off scales \cite{Banfi:2012jm}.
Hence, this cut was optimized by looking for both the largest $S/\sqrt{B}$ and the largest jet \pT\ threshold. The optimal threshold was found to be 75\,GeV.
Finally, events must have at least one jet with \pT\ $\geq$ 35\,GeV and $2.5 < |\eta| < 4.5$, consistent with the single top partner production mechanism.

In Table~\ref{tab:CutFlow}, the number of expected events for 20\,fb$^{-1}$ integrated luminosity at the 8\,TeV and 14\,TeV LHC is shown at the preselection stage, after requiring the central jet-veto, and  after the final selection. The selection cuts lead to a $S/B$ of 0.22 (0.20) and $S/\sqrt{B}$ of 2.5 (5.65) for a 700\,GeV $\tilde{T}$ at 8\,TeV(14\,TeV). The addition of the forward jet requirement leads to 
modest improvements in $S/\sqrt{B}$ but more than doubles $S/B$, it can thus be very useful in the case of large experimental uncertainties on the background normalization.

To search for $\tilde{T}$ production, a candidate $\tilde{T}$-mass constructed from the charged lepton and the $b$-jet
and neutrino candidates is used . The $b$-tagged small-radius jet, that is matched to the large radius jet, is used as the $b$-jet candidate. The neutrino candidate is constructed from the \MET\ and charged lepton by imposing a $W$-mass constraint on the lepton+ \MET\ system~\cite{Aad:2013nca}. 
The resulting $\tilde{T}$-candidate mass distribution is shown for signal and backgrounds
in Figure~\ref{fig:Tmass} for 20\,fb$^{-1}$ integrated luminosity at the 8\,TeV LHC. 
The signal distributions peak close to the $\tilde{T}$ mass while background distributions turn over at lower masses.

\bgroup
\def\arraystretch{1.2}
\begin{table*}[!t]
\begin{tabular}{|c|c|c|c|c|c|c|}
\hline
& \multicolumn{6}{c|} {Number of events} \\
\cline{2-7}
& \multicolumn{2}{c|} {after preselection} & \multicolumn{2}{c|} {after jet veto} & \multicolumn{2}{c|} {after final selection} \\
\cline{2-7}
& 8\,TeV  & 14\,TeV  & 8\,TeV  & 14\,TeV & 8\,TeV  & 14\,TeV \\ 
\hline
$W$+jets                                     &  40.1$\times 10^{6}$ & 87.3$\times 10^{6}$     &  391      &  1403.2    &  92   &  368.1 \\ 
$t\bar{t}$                                     &  1.13$\times 10^{6}$ &    4.1$\times 10^{6}$    &  222.1   &    887    & 32   &   217.7\\ 
Single top                                     &          2176.8             & 25981                           &      28    &   380     & 13   &  223.2  \\ 
\hline
Total Background                         & 41.3$\times 10^{6}$  & 91.4$\times 10^{6}$     & 640.6    &   2670       & 137  &  809   \\ 
 \hline
Signal ($m(\tilde{T})=700$\,GeV) & 664.5                        &  3653                             &   54.4   &   234.3       &   30  & 161   \\ 
 \hline
\hline
$S/B$                                           & 1.61$\times 10^{-5}$  & 4$\times 10^{6}$         &   0.085  & 0.088        & 0.22 & 0.20    \\ 
$S/\sqrt{B}$                                 & 0.1                              & 0.38                             &  2.15     &  4.53      & 2.5   & 5.65   \\ 
\hline
\end{tabular}
\caption{The number of expected events for an integrated luminosity of 20\,fb$^{-1}$ at the LHC with 8\, and 14\,TeV.}
\label{tab:CutFlow}
\end{table*}

A comparison was made with the strategy presented in Ref.~\cite{Li:2013xba}, where the authors address the $Wb$ final state by using a high \pT\ lepton, large summed transverse momentum (H$_{\mathrm{T}}$) and the masses 
                 of combinations of final state particles.
                 After running both selections on two signal samples with masses of 700 and 800 \,GeV,
                 it was found that although the $S/\sqrt(B)$ for both analyses is comparable,  our approach 
                 accepts between 50\% and 70\% more events.
This will either allow for additional cuts to purify the final state or in an improved reach of the resonance mass or smaller production cross section.

\section{Results}

\begin{figure*}[htb]
  \includegraphics[width=0.43\textwidth]{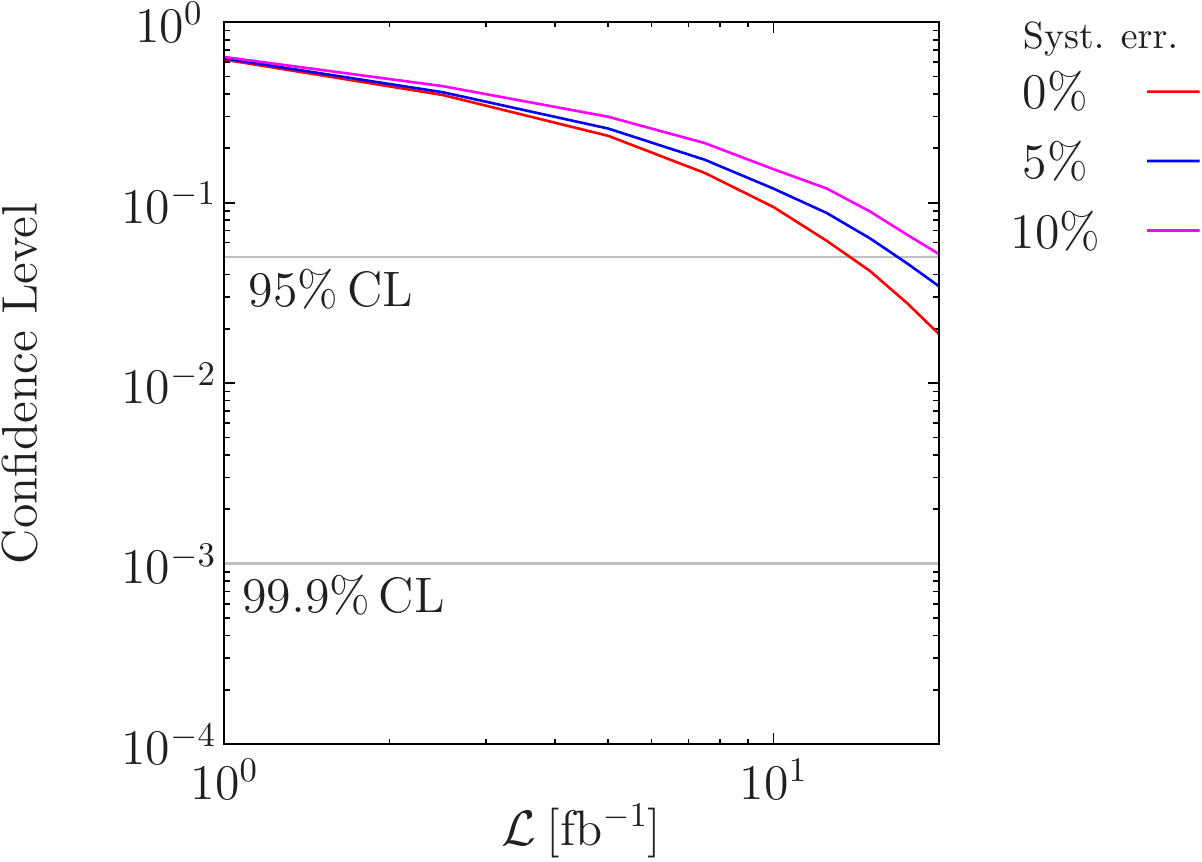}
  \hspace{0.05\textwidth}
  \includegraphics[width=0.43\textwidth]{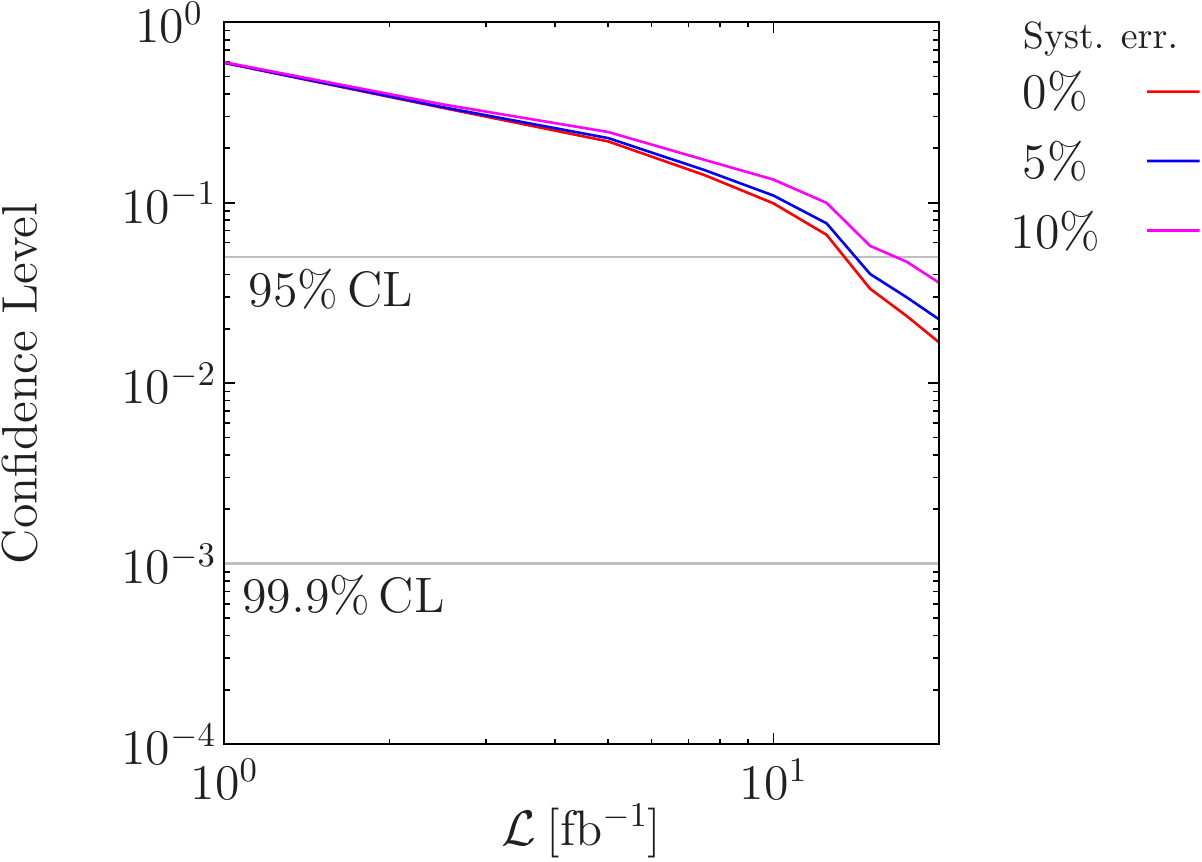}
  \includegraphics[width=0.43\textwidth]{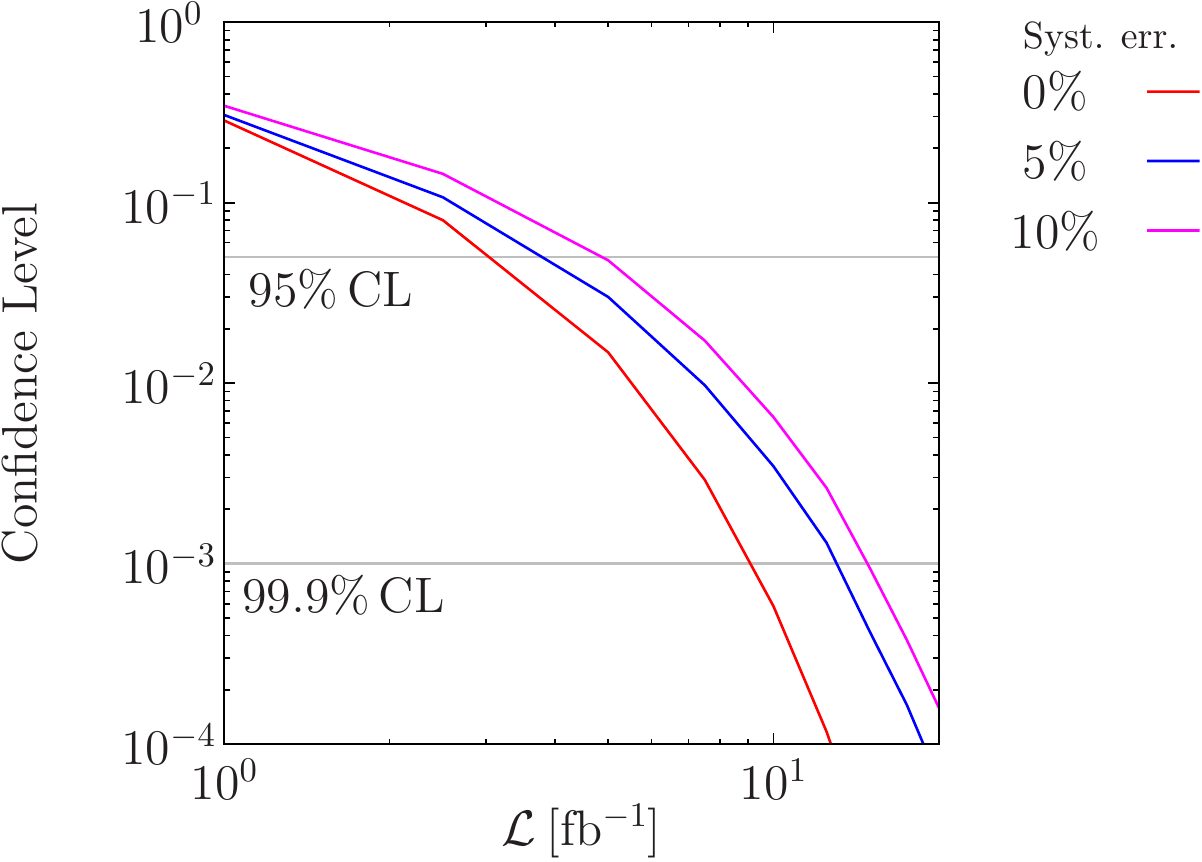}
  \hspace{0.05\textwidth}
  \includegraphics[width=0.43\textwidth]{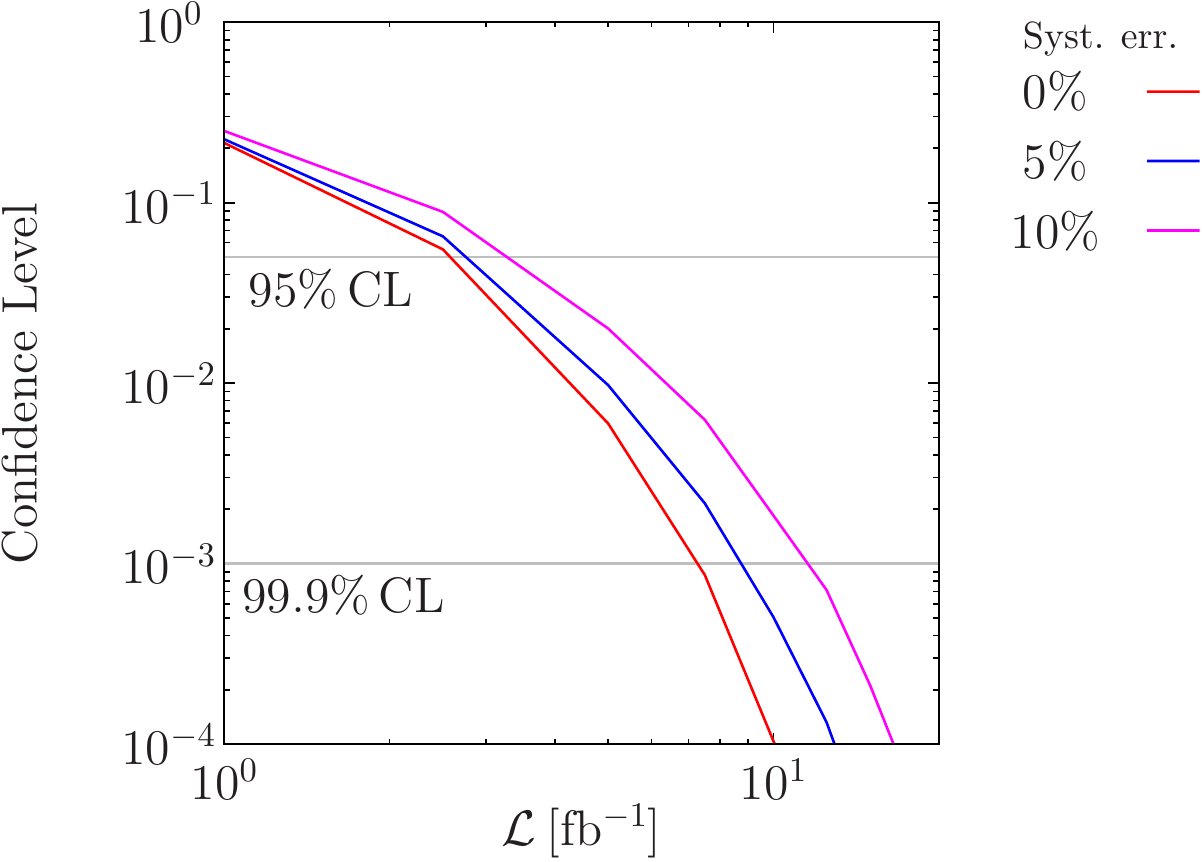}
  \caption{The results of the limit-setting procedure when applied to the search for
a 700\,GeV $\tilde{T}$ using 8\,TeV (top) and 14\,TeV (bottom) data. The results without requiring a forward jet (left) can be directly compared to the results with a forward jet requirement (right).}
  \label{fig:CLs700at8} 
\end{figure*}

In order to estimate whether a singly-produced top partner can be excluded with data
 collected at the LHC, we use a binned log-likelihood hypothesis test as 
described in Ref. \cite{Junk:1999kv}. The results of this procedure, as a function of integrated luminosity for 8\,TeV and 14\,TeV LHC data, are presented in 
Figure~\ref{fig:CLs700at8}, which shows the confidence level at which single
 production of a 700\,GeV  $\tilde{T}$ quark can be excluded in the absence of signal. These results are shown both before and after the forward-jet requirement and for different assumed levels of systematic uncertainty on the backgrounds: 0\%, 5\% and 10\%. It can be seen that the sensitivity is improved by the forward-jet requirement and that the full selection is sufficient to
exclude the single-top partner production using 8\,TeV data even for the most pessimistic assumed systematic uncertainty. Although this systematic treatment is rather simplistic
we expect the search strategy to retain sufficient sensitivity to exclude the
700\,GeV $\tilde{T}$ if deployed in a real experimental analysis. For 14\,TeV data, the exclusion of the full 8\,TeV data set is already surpassed at an integrated luminosity of 5\,$\mathrm{fb^{-1}}$.

\section{Summary, Conclusions and Outlook}

This is the first study to demonstrate a simple and feasible strategy for discovering 
single production of a top partner in the $\tilde{T} \rightarrow Wb$ channel.
The study shows that use of a forward jet and the requirement of a low-mass central large-radius jet and central jet veto are powerful tools for rejecting the SM background processes whilst retaining acceptable signal efficiency.
This analysis is indicative that single production of 700\,GeV top partners could be excluded already at the 8\,TeV LHC. The mass reach of searches at higher-energy LHC running is likely to be significantly extended.  This study considers $\tilde{T}$ production within a specific composite Higgs scenario. However,  the analysis is generic and its applicability to top partner searches in other models can be inferred from the production cross sections in our chosen model and from the exclusion limits shown in Figure.~\ref{fig:CLs700at8}. We reserve study of 
$\tilde{T} \rightarrow tH$ and $\tilde{T} \rightarrow tZ$ for a future work, 
noting that in most searches the extra efficiency in a dominant decay channel 
outweigh signal purity in the selected sample.

{\emph{Acknowledgements.}} 
We thank Andy Buckley and Karl Nordstrom for assistance with Rivet, Frank Siegert for advice on \sherpa configuration, and Andrew Pickford for computing support. We thank Oleskii Matsedonskyi, Andrea De Simone and Andrea Wulzer for providing valid parameter points for the composite Higgs scenarios under consideration. We thank Tony Doyle for useful discussion.
 The work of JF and DK is supported by the UK Science and Technology Facilities Council (STFC). The work of NG is supported by the Scottish Universities Physics Alliance (SUPA).

%


\end{document}